\documentclass{IEEEtran4PSCC}
\usepackage[dvipsnames]{xcolor}
\usepackage{multirow}
\usepackage{cite}

\ifCLASSINFOpdf
   \usepackage[pdftex]{graphicx}
\else
   \usepackage[dvips]{graphicx}
\fi
\usepackage[cmex10]{amsmath}
\usepackage{amssymb}
\usepackage{mathtools}
\usepackage[nodisplayskipstretch]{setspace}

\usepackage{array}

\usepackage{xcolor}
\usepackage{soul}

\makeatletter
\let\old@ps@headings\ps@headings
\let\old@ps@IEEEtitlepagestyle\ps@IEEEtitlepagestyle
\def\psccfooter#1{%
    \def\ps@headings{%
        \old@ps@headings%
        \def\@oddfoot{\strut\hfill#1\hfill\strut}%
        \def\@evenfoot{\strut\hfill#1\hfill\strut}%
    }%
    \def\ps@IEEEtitlepagestyle{%
        \old@ps@IEEEtitlepagestyle%
        \def\@oddfoot{\strut\hfill#1\hfill\strut}%
        \def\@evenfoot{\strut\hfill#1\hfill\strut}%
    }%
    \ps@headings%
}
\makeatother

\begin{document}

\title{Representative Days and Hours with Piecewise Linear Transitions for Power System Planning}

\author{
\IEEEauthorblockN{Mojtaba Moradi-Sepahvand}
\IEEEauthorblockA{Department of Electrical Sustainable Energy \\
Delft University of Technology\\
Delft, The Netherlands\\
\ m.moradisepahvand@tudelft.nl}
\and
\IEEEauthorblockN{Simon H.~Tindemans}
\IEEEauthorblockA{Department of Electrical Sustainable Energy \\
Delft University of Technology\\
Delft, The Netherlands\\
\ s.h.tindemans@tudelft.nl}
}

\maketitle

\begin{abstract}
Electric demand and renewable power are highly variable, and the solution of a planning model relies on capturing this variability. This paper proposes a hybrid multi-area method that effectively captures both the intraday and interday chronology of real data considering extreme values, using a limited number of representative days, and time points within each day. An optimization-based representative extraction method is proposed to improve intraday chronology capturing. It ensures higher precision in preserving data chronology and extreme values than hierarchical clustering methods. The proposed method is based on a piecewise linear demand and supply representation, which reduces approximation errors compared to the traditional piecewise constant formulation. Additionally, sequentially linked day blocks with identical representatives, created through a mapping process, are employed for interday chronology capturing. To evaluate the efficiency of the proposed method, a comprehensive expansion co-planning model is developed, including transmission lines, energy storage systems, and wind farms.    
\end{abstract}


\thanksto{\noindent This work is part of the project Heuristic Efficient Proxy-based Planning of Integrated Energy Systems (HEPPIE), funded by R\'eseau de Transport d’\'Electricit\'e (RTE).}

\section{Introduction}
In modern power systems, capturing variability in electrical load and renewable energy source (RES) outputs during operational periods is vital for effective power system planning. However, integrating short-term operational variability into long-term studies presents a substantial challenge. Attempting to account for every hourly time period across multiple years results in large intractable problems. Therefore, to accurately represent the variability in load and RES output power, the extraction of suitable representative time periods becomes necessary. Commonly employed in the literature \cite{moradicapturing,poncelet2016selecting,gao2021spectral,li2022representative}, extracting representative days (RDs) using clustering algorithms has been a prevailing method. These RDs, each spanning 24 hours, must faithfully maintain the chronology and extreme values of the underlying time series. Notably, when long-term energy storage systems (ESSs) are involved, interday chronology preservation becomes essential \cite{moradicapturing}.

Past research, such as \cite{poncelet2016selecting}, employed hierarchical clustering to extract typical RDs for weekends, weekdays, and holidays in each season. Spectral clustering, as presented in \cite{gao2021spectral}, was utilized for RD extraction by considering net load and the ramp of net load duration curves. However, both approaches lacked consideration for interday chronology and capturing extreme values. The study by \cite{li2022representative} introduced an RD selection method focused on capturing extreme operating conditions for power system planning while disregarding interday chronology. Additionally, \cite{orgaz2022modeling} employed a time series aggregation (TSA) method, relying on system state extraction, to maintain chronological data sequence while disregarding extreme values. Notably, the considerable number of sequential hourly system states in \cite{orgaz2022modeling} does not significantly decrease the problem complexity \cite{gonzato2021long}.

Efforts to capture data chronology introduced the chronological time period clustering (CTPC) method in \cite{pineda2018chronological}. In addition, this method's inability to preserve extreme values was addressed and improved in \cite{garcia2022priority}. However, CTPC-based methods remained inadequate for long-term ESS cycle modeling \cite{moradicapturing}. To address these challenges, \cite{moradicapturing} developed a hybrid clustering-based algorithm. It effectively maintained chronology and extreme values by introducing sequentially linked day blocks (SLDs) for long-term ESS modeling. According to comprehensive reviews of TSA approaches presented in \cite{teichgraeber2022time,gonzato2021long}, many papers employed RDs with 24-hour periods, applying the same data patterns for all system buses \cite{moradicapturing,poncelet2016selecting,gao2021spectral,li2022representative,orgaz2022modeling,pineda2018chronological,garcia2022priority}. Despite this, further complexity reduction can be achieved by extracting appropriate representative hours (RHs) or time points (RTPs) within each RD. The complexity challenge is highlighted by the fact that power systems are diverse, exhibiting distinct load and RES generation patterns across multiple areas. Consequently, adopting identical load and RES generation patterns for all buses yields unrealistic results. 

Conventional power system optimal power flow (OPF) models implicitly utilize piecewise constant (PWC) formulations, describing injections and loads with average power levels for a time interval (effectively energy). However, PWC models tend to overestimate flexibility, as they inaccurately model ramping and reserves \cite{nycander2022power}. In contrast, piecewise linear (PWL) OPF formulations (also called `power-based') offer improved operational flexibility by more accurately representing instantaneous power trajectories \cite{tejada2019power,nycander2021security}. In \cite{nycander2022power}, a power-based (i.e., PWL) economic dispatch model is introduced, and \cite{nycander2021security} presents a security-constrained unit commitment power-based formulation, both offering superior accuracy compared to traditional energy-based models. Similarly, the power-based generation expansion planning model in \cite{tejada2019power} demonstrates enhanced accuracy in representing flexibility requirements while considering short-term ESSs. However, this model neglects long-term ESS and multi-area data considerations, as well as inter-period chronology and the preservation of extreme values. 

To tackle these limitations, this paper introduces a multi-area method that effectively captures intraday and interday chronology, while considering extreme values. The proposed method achieves this using a limited number of RDs and RTPs within each RD, as well as an adaptive optimization-based procedure to improve intraday chronology and extreme values capturing by allocating RTPs across RDs in accordance with the complexity of each day. Furthermore, in order to capture the interday chronology necessary for modeling long-term ESSs, the proposed method builds on the ERD methodology \cite{gonzato2021long} in combination with the SLD process to merge adjacent days introduced in \cite{moradicapturing}. 
We further enhance our method by transforming the problem formulation from a traditional PWC to a PWL formulation. This enhancement contributes to improved modeling accuracy by better representing operational flexibility. To validate the proposed method effectiveness, an expansion co-planning model is developed that includes transmission lines, both short (battery) and long-term (hydropower pumped) ESSs, and wind farms. In what follows, the main contributions are outlined:
\begin{itemize}
    \item Proposing an optimization-based PWL RTP extraction method to improve intraday chronology and extreme value capturing by extracting both equal and unequal numbers of RTPs within each RD.
    \item Improving generation cost modeling for thermal units within the PWL framework. 
    \item Developing a multi-area PWL-adapted clustering-based algorithm for extreme-value sensitive RD extraction.    
\end{itemize}

\section{Reduction of Days and Time Points}

\subsection{Piecewise Linear Model}
In this paper, we use PWL OPF formulations in the proposed co-planning model. This offers improved modeling of ramping and spinning reserve in comparison with traditional PWC models. More importantly, it is anticipated that these benefits \emph{grow} as the model representation becomes increasingly coarse-grained, i.e., when fewer RTPs are used. 
 
The shift from PWC to PWL necessitates a careful description of time in an optimization model. In the PWC representation, measurements refer to an \emph{interval} of time, whereas in the PWL representation, values refer to an \emph{instant} in time. 

In the following, the index $t$ is an integer time coordinate that refers to a point in time, and a pair $(t-1,t)$ refers to a time interval. 
To give a specific example of the PWL interpretation, snapshot power measurements $P_t$ can be thought to represent a \emph{continuous} function of $t'$:
\begin{equation} \label{eq:linearpower}
    P(t') = P_t + (t' - t) (P_{t+1} - P_t) \qquad \text{for   } t' \in [t,t+1]
\end{equation}
The energy consumed/produced during the interval $[t,t+1]$ is
\begin{equation}
    E(P_t, P_{t+1}) = \tfrac12 \Delta_t (P_t + P_{t+1})
\end{equation}
where $\Delta_t$ is the (real) time between time points indexed by $t$ and $t+1$.

\subsection{Representative Days}
In this paper, the extreme-value sensitive method in \cite{moradicapturing} is developed as a multi-area clustering-based algorithm to select RDs. Input data consisting of hourly measurements of multiple features (wind, load) in multiple areas, was first divided into daily sequences consisting of 25 hourly measurements (with midnight being represented in both adjacent days). Euclidean distance between days (incl.~all features and areas) was used to hierarchically cluster days until the target number was reached. By default, days during which one area's maximum net load (load minus maximum available wind capacity) occurs, are marked as \emph{extreme days}. These days are preserved, to improve planning for system adequacy. 

After extracting RDs, the SLDs are created through a mapping process according to \cite{moradicapturing}. The outputs of the developed RD extraction method are: sets of unordered RDs, i.e., $\mathcal{D}$, and ordered SLDs, i.e., $\mathcal{SD}$, the weight of each RD $d$, i.e., $\omega_d$, the association matrix $D_{sd,d}$ between RD $d$ and SLD $sd$, the number $n^{B}_{sd}$ of RD repetitions within an SLD $sd$, and instant load and wind representative factors in each area $a$, RD $d$ and time $t$, i.e., $FL_{a,d,t}$ and $FW_{a,d,t}$.

\subsection{Sparse Day Representation}
In the PWL modeling framework, a representative day is initially represented by 25 hourly time points (midnight to midnight). In this section, we describe an optimization-based method to reduce this number with minimum loss of modeling accuracy.
Two variations of the method are proposed: one in which a specified number of time points is extracted from a single RD, and another in which a total number of time points is specified for all RDs. The latter 
makes it possible to extract unequal numbers of RTPs within each RD, allowing for the extraction of more RTPs from RDs with higher hourly variation. 
The optimization for a single day is given by
\begin{subequations} \label{eq:opt_for_RTP_all}
\begin{align}
& {O}\left(y_{a,t,f}^{RD},\bar{r}\right)=\quad \min_{\mathclap{Z_{a,t,f},I_t,ER_{a,t,f}^{+,-}}} \qquad \textstyle \sum_{a,t,f} ER_{a,t,f}^{+}+ER_{a,t,f}^{-} \label{eq:opt_for_RTP_a} \\ 
& ER_{a,t,f}^{+}-ER_{a,t,f}^{-}=y_{a,t,f}^{RD}-Z_{a,t,f} \hspace{0.8 cm} \forall a,t,f \label{eq:opt_for_RTP_c} \\
& ER_{a,t,f}^{+} \geq 0, \quad ER_{a,t,f}^{-} \geq 0, 
\hspace{1.8 cm} 
\forall a,t,f \label{eq:opt_for_RTP_h}\\ & I_t \in\{0,1\} \hspace{4.53cm} \forall t  \label{eq:opt_for_RTP_i} \\
& \textstyle \sum_t I_t=\bar{r}, \quad \quad I_0=I_{24}=1   \label{eq:opt_for_RTP_b}\\
& Z_{a,t,f} \leq y_{a,t,f}^{RD}+ M \left(1-I_t\right) \hspace{2.25 cm} \forall a,t,f \label{eq:opt_for_RTP_d}\\
& Z_{a,t,f} \geq y_{a,t,f}^{RD}-M \left(1-I_t\right) \hspace{2.25 cm} \forall a,t,f \label{eq:opt_for_RTP_e}\\
& Z_{a,t,f} \leq 1/2\left(Z_{a,t-1,f}+Z_{a,t+1,f}\right)+M I_t \quad \forall a,t,f \label{eq:opt_for_RTP_f}\\
& Z_{a,t,f} \geq 1/2\left(Z_{a,t-1,f}+Z_{a,t+1,f}\right)-M I_t \quad \forall a,t,f \label{eq:opt_for_RTP_g}
\end{align}
\end{subequations}
Here $a$, $t$, and $f$ are indices of area, time points, and feature vectors. $y_{a,t,f}^{RD}$ are the values of the relevant feature vector (load, wind) for the given RD $d$. For $f=\mathrm{wind}, y_{a,t,f}^{{RD}}=FW_{a,d,t}$; for $f=\mathrm{load}, y_{a,t,f}^{{RD}}=FL_{a,d,t}$. The objective \eqref{eq:opt_for_RTP_a} is to minimize the integrated absolute error between $y_{a,t,f}^{RD}$ and its approximation $Z_{a,t,f}$, summed over all areas, time points, and features. Positive and negative deviations are accumulated in $ER_{a,t,f}^{+}$ and $ER_{a,t,f}^{-}$, respectively (\eqref{eq:opt_for_RTP_c}-\eqref{eq:opt_for_RTP_h}). The sparse representation is constructed by selecting $\bar{r}$ time points using binary variables $I_t$, where the first and last time of the day are always selected \eqref{eq:opt_for_RTP_b}.
Constraints \eqref{eq:opt_for_RTP_d}-\eqref{eq:opt_for_RTP_e} indicate that if a time point is selected (i.e., $I_{t}=1$), $Z_{a,t,f}$ takes the value of $y_{a,t,f}^{RD}$ (a big-$M$ formulation is used). Otherwise, it takes the average of the neighboring values 
(\eqref{eq:opt_for_RTP_f}-\eqref{eq:opt_for_RTP_g}; this is not required for $t=0$ and $t=24$).

\begin{figure}[!b]
\centering
\includegraphics[scale=0.3466]{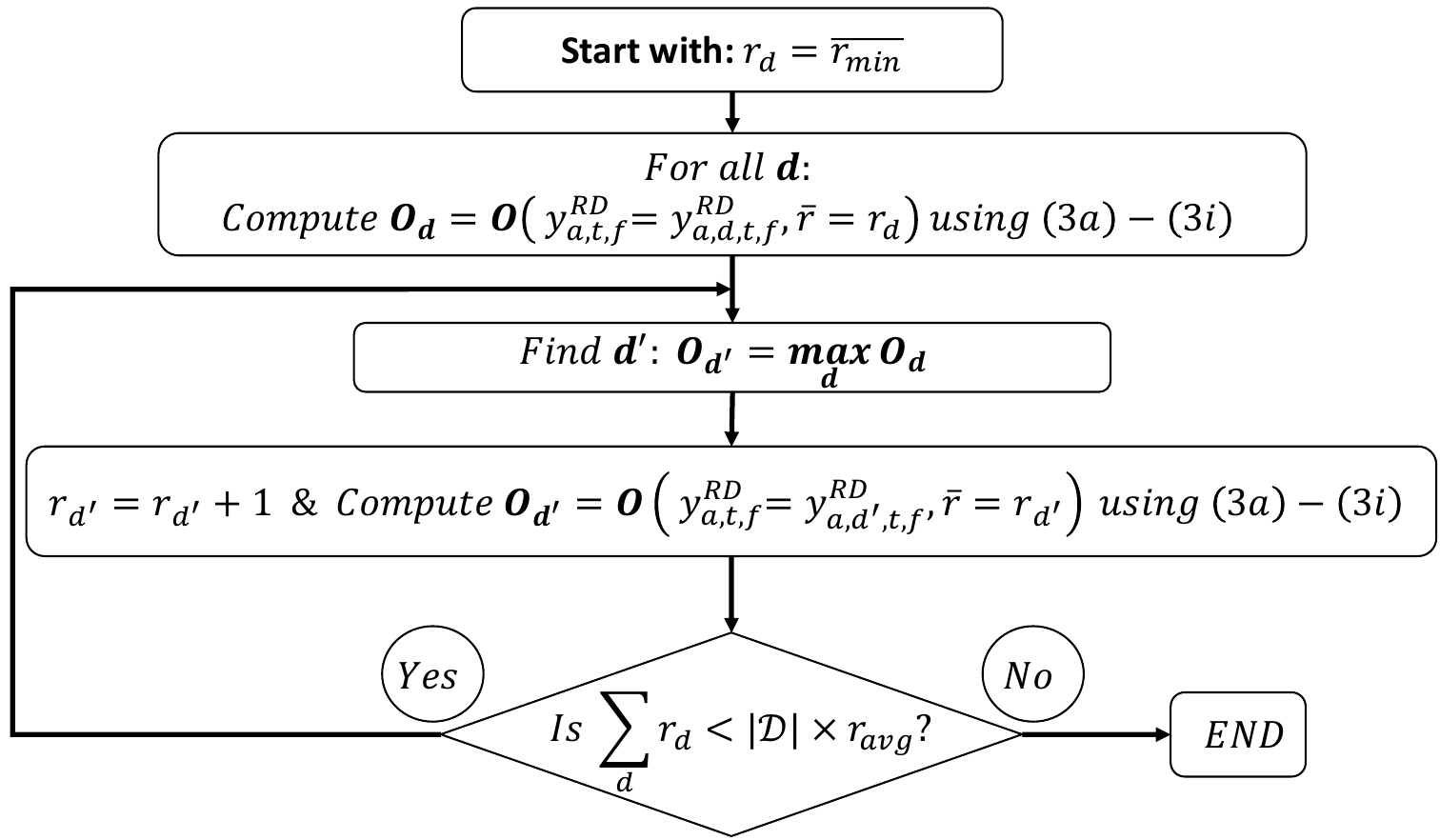}
\caption{Proposed optimization-based algorithm to extract RTPs from each RD} 
\label{fig:01}
\end{figure}

The algorithm to balance RTPs across RDs minimizes the maximum error across RDs using a greedy algorithm. A  flowchart for the algorithm is depicted in Fig. \ref{fig:01}. First, a minimum number of RTPs ($r_d=\bar{r}_{min}$) is assigned to each RD. After solving the optimization \eqref{eq:opt_for_RTP_all} for all RDs, the day with the largest mismatch is identified and the number of RTPs for that day is increased by one. The algorithm continues until the total number of RTPs is equal to the total number of RDs ($|\mathcal{D}|$) times the desired average number of RTPs ($r_{avg}$) per RD. 

After termination of the RTP selection algorithm, only the $r_d$ selected RTPs within each day are extracted for further use. Let $J_{d,k}$ be the index (hour) of the $k$-th selected RTP on day $d$, with $J_{d,0}=0$ and $J_{d,r_d-1}=24$. For consistency of notation between models, we let index $t$ refer to the selected RTPs, so e.g., $FW_{a,d,t} \leftarrow FW_{a,d,J_{d,t}}$. Moreover, to account for unequal time steps, we define 
\begin{equation}
\Delta_{d,t} = J_{d,t+1}-J_{d,t}, \hspace{1 cm}  \forall d,t\in{\{0,\dots,r_d-1}\} \label{durationRTP_2}
\end{equation}

\section{Co-planning model formulation}
This section describes the proposed PWL co-planning model, both in its reference form (all days, all time points) and variations with limited RDs and RTPs. 

\subsection{Model Choices and Notation}

This formulation is based on the notion of \emph{areas} that consist of one or more electrical \emph{buses} (indexed by $b$). The allocation of buses to areas is represented by the matrix $M^B$, where $M^B_{b,a}$ is 1 if bus $b$ is located in area $a$, and 0 otherwise. Similarly, the assignment of load demand $l$, thermal generators $g$, wind power generators $w$ and storage systems $s$ to buses $b$ is given by matrices $M^L$, $M^G$, $M^W$ and $M^S$, respectively, with $M^X_{b,x}$ indicating the assignment of resource $x$ to bus $b$. 

Iterations or additions over sets are commonplace in optimization problems. For notational simplicity, $\forall x$, where $x$ is a relevant index, is used as shorthand for $\forall x \in \mathcal{X}$ throughout this section: the relevant set or tuple is implicit. Indices $b,a,l,g,w,s$ have corresponding sets $\mathcal{B}, \mathcal{A}$, $\mathcal{L}, \mathcal{G}, \mathcal{W}$ and $\mathcal{S}$, respectively. Time points $t$ are part of the ordered tuple $\mathcal{T}$, and whenever two neighboring time points (e.g., $t$ and $t+1$) are referenced, $\forall t$ only refers to those $t$ for which both are valid indices. Days $d$ may be part of an ordered tuple (the reference case) or unordered set (RDs) $\mathcal{D}$. Indices $sd$ are elements of the ordered set $\mathcal{SD}$. 

Wind power is the only renewable resource in the present formulation, but other variable power resources such as solar PV can be added in the same way. The ESSs are implemented as long-term units that require state variables to track an annual state of charge. More generally, the presentation in this section prioritizes clarity and compactness over the maximum efficiency of resulting equations. Generic equations for investment and operation are presented in sections \ref{sec:investment} and \ref{sec:operations}, followed by three specializations: reference model, RD model, and RDTP (RD and RTP) model. Some redundant equations could be removed for each of the resulting models.

\subsection{Investment Options} \label{sec:investment}
Investment options in the proposed co-planning model are new transmission lines, ESSs, and wind farms. The corresponding total cost of investment (\emph{CI}) is given by
\begin{equation}
\begin{aligned} 
CI &= \textstyle \sum_{nl} \left[CL_{nl} LL_{nl} Y_{nl}\right] \\ &+\textstyle \sum_{s} \left[CE_s E^{cap}_{s} +CC_s C^{cap}_{s}\right]+\textstyle \sum_{w} \left[CW_w W_{w}\right] \label{eq:investmentcost}
\end{aligned}
\end{equation}
All investment costs are converted to equivalent annual costs using the capital recovery factor of each investment option \cite{moradi2023secure}. ${CL_{nl}, LL_{nl}}$ and ${Y_{nl}}$ are investment cost (\$/km), length (km), and construction binary variable of new lines, respectively. ${E^{cap}_s, C^{cap}_s, CE_s}$ and ${CC_s}$ are ESS energy and power capacity variables and related investment costs, respectively. ${W_w}$ and ${CW_w}$ are the power capacity of the installed wind farm and its investment cost. The investment constraints are given by
\begin{subequations} 
\label{eq:invcon}
\begin{align}
\label{eq:ESSinvconst} E^{cap}_s & \le E^{max}_s, \ C^{cap}_s \le C^{max}_s, \ C^{cap}_s \phi_s \le E^{cap}_s && \forall s \\ 
\label{eq:WFinvconst1} 0 \le & W_w \le W^{max}_w && \forall w \\
\label{eq:WFinvconst2} 25\% & \textstyle\sum_{l} L_l \le \textstyle\sum_{w} W_w
\end{align}
\end{subequations}
here, $L_l$ is the peak load in each load bus. Energy and power capacity along with the energy-to-power ratio for each ESS is bounded in \eqref{eq:ESSinvconst}. $\phi_s$(h) is the energy (MWh) to power (MW) ratio for each ESS. The capacity of wind farms, considering a renewable portfolio standard policy is limited by \eqref{eq:WFinvconst1} and \eqref{eq:WFinvconst2}.

\subsection{Operational Model} \label{sec:operations}
In this subsection objective function and all related constraints for modeling operational details in reference form and variations with RDs and RTPs are presented.

\subsubsection{Operational Costs}
The cost of operation for the day $d$ is given by
\begin{equation}
    CO_{d} =\textstyle \sum_{t} \left[\textstyle \sum_{g}  CG_{g,d,t} +\textstyle \sum_{l} CS_{l,d,t} \right]
\end{equation}
where $CG_{g,d,t}$ and $CS_{l,d,t}$ are the cost of thermal generation and involuntary load shedding in the interval $[t,t+1)$, respectively. Note that we do not consider fixed operational costs in the current model, but relevant terms can readily be added. The cost of load shedding is modeled as
\begin{align} \label{LScost}
    CS_{l,d,t} & = \frac{VOLL }{2} (PS_{l,d,t} + PS_{l,d,t+1})\Delta_{d,t} && \forall l,d,t 
\end{align}
where $VOLL$ is the value of lost and $PS_{l,d,t}$ is the lost load. In this paper, thermal generation cost is modeled with a quadratic instantaneous cost curve
\begin{equation}
    cg_g(P) = \tfrac12 a_g P^2 + b_g P \nonumber
\end{equation}
This quadratic cost function, combined with the generic PWL curve \eqref{eq:linearpower}, results in the following cost for the interval $[t,t+1)$ with duration $\Delta_t$:
\begin{align}
    cg_g(P_t, P_{t+1}) &= \Delta_t \int_0^1 \left[ cg_g(P(t')) \right] \mathrm{d} t' \nonumber \\
    &= \Delta_t  cg_g(\tfrac12(P_t+P_{t+1})) + \frac{a_g \Delta_t}{24}(P_t-P_{t+1})^2  \nonumber
\end{align}
where the result is followed by integration and collection of terms. We approximate the thermal generation costs $CG_{g,d,t}$ by the linear lower convex envelope determined by bounding each of the two terms:
\begin{subequations}
\begin{align} \label{eq:costfunction}
    CG_{g,d,t} & = CG^{(1)}_{g,d,t} + CG^{(2)}_{g,d,t} \hspace{1.8cm} \forall g,d,t \\
    \frac{CG^{(1)}_{g,d,t}}{\Delta_{d,t}} & \ge   (a_g \pi_{g,k} + b_g) 
    \frac{PG_{g,d,t} + PG_{g,d,t+1}}{2} - \tfrac12 a_g \pi_{g,k}^2  \nonumber \\
    & \hspace{5cm}\forall g,k,d,t \label{eq:linconstraint1} \\
    \frac{CG^{(2)}_{g,d,t}}{\Delta_{d,t}} & \ge   \frac{a_g \pi_{g,k}}{12} (PG_{g,d,t} - PG_{g,d,t+1})  - \tfrac{1}{24} a_g \pi_{g,k}^2   \nonumber \\
    & \hspace{5cm} \forall g,k,d,t \label{eq:linconstraint2} \\     
    \frac{CG^{(2)}_{g,d,t}}{\Delta_{d,t}} & \ge   -\frac{a_g \pi_{g,k}}{12} (PG_{g,d,t} - PG_{g,d,t+1})  - \tfrac{1}{24} a_g \pi_{g,k}^2  \nonumber \\
    & \hspace{5cm} \forall g,k,d,t \label{eq:linconstraint3} \\   
    \pi_{g,k} &= \left(\tfrac{k-1}{K-1} \right) PG^{max}_{g} \hspace{2.3cm} \forall g,k 
\end{align}
\end{subequations}
where $k \in (1,\dots,K)$ indexes the tangent lines and $\pm \pi_{g,k}$ are the power levels for which the linear constraints \eqref{eq:linconstraint1}-\eqref{eq:linconstraint3} are binding. Note that $PG_{g,d,t}$ is the thermal generator output power.

\subsubsection{Power Balance}
The local power balance equations enforce the conservation of power in every bus. 
\begin{multline}
    \textstyle \sum_l M^L_{b,l}PL_{l,d,t} + \sum_s M^S_{b,s} PC_{s,d,t} = \\
    \textstyle  \sum_g M^G_{b,g}PG_{g,d,t} + \sum_w M^W_{b,w}PW_{w,d,t}  + \sum_s M^S_{b,s} PD_{s,d,t} \\
   \textstyle + \sum_l M^L_{b,l}PS_{l,d,t}  +  PN_{b,d,t} \qquad \forall b, d,t
\end{multline}
here, $PL_{l,d,t}$ and $PW_{w,d,t}$ are bus-specific demand pattern and wind farm dispatched power. $PC_{s,d,t}$ and $PD_{s,d,t}$ are charging and discharging power of each ESS. $PN_{b,d,t}$ denotes the power transported to node $b$ by the network. The network consists of lines $el \in \mathcal{EL}$ that are present throughout and lines $nl \in \mathcal{NL}$ that are optionally constructed, the presence of which is indicated by the binary variable $Y_{nl}$. The location and orientation of lines are determined by the \emph{directed} bus-line incidence matrices $A^{EL}_{b,el}, A^{NL}_{b,nl}$ and the oriented power flows are given by the variables $FE_{el,d,t}, FN_{nl,d,t}$. Then, the flows are determined by the DC power flow equations given by
\begin{subequations}
\begin{align}
PN_{b,d,t}&= \textstyle \sum_{el} A^{EL}_{b,el} FE_{el,d,t} \nonumber \\ &+ \textstyle \sum_{nl} A^{NL}_{b,nl} FN_{nl,d,t} \hspace{1.05cm} \forall b,d,t \\
FE_{el,d,t} & = B_{el} \textstyle \sum_b A^{EL}_{b,el}  \theta_{b,d,t} \hspace{1.15cm} \forall el,d,t \\
 -FE_{el}^{max} & \le  FE_{el,d,t}  \le FE_{el}^{max} \hspace{1cm} \forall el,d,t \\
  -(1-Y_{nl}) M_{nl} & \le FN_{nl,d,t} - B_{nl} \textstyle \sum_b A^{NL}_{b,nl}  \theta_{b,d,t} \nonumber \\ \le (1-Y_{nl}) M_{nl} & \hspace{4.35cm} \forall nl,d,t \\
 -FN_{nl}^{max} Y_{nl} & \le  FN_{nl,d,t}  \le FN_{nl}^{max} Y_{nl} \hspace{0.5cm} \forall nl,d,t
\end{align}
\end{subequations}
where $\theta_{b,d,t}$ is the voltage phase angle, $M_{nl}$ is big-M for each \emph{nl}, and $B_{el(nl)}$ is the line susceptance.

\subsubsection{Generation Dispatch and Curtailment}
Thermal generators are constrained by the power and ramp constraints
\begin{subequations}
\begin{align}
    0 \le PR_{g,d,t} & \hspace{3.29cm} \forall g,d,t \\
    PR_{g,d,t} \le PG_{g,d,t} & \le PG^{max}_g - PR_{g,d,t} \hspace{0.18cm} \forall g,d,t\\
    \left|\frac{PG_{g,d,t+1} - PG_{g,d,t} }{\Delta_{d,t}}\right| & + \frac{PR_{g,d,t}}{\tau} \le R^{max}_g \hspace{0.44cm} \forall g,d,t \label{eq:starthour} \\
    \left|\frac{PG_{g,d,t+1} - PG_{g,d,t} }{\Delta_{d,t}}\right| & + \frac{PR_{g,d,t+1}}{\tau} \le  R^{max}_g \hspace{0.16cm} \forall g,d,t \label{eq:endhour}
\end{align}
\end{subequations}
where $PR_{g,d,t}$ is the flexible spinning reserve requirement (identical up and down) allocated to the generator $g$ and $\tau < \Delta_{d,t}$ is its delivery time, and $R^{max}_g$ is the maximum ramp rate (also up and down). 
Eqs.~\eqref{eq:starthour}-\eqref{eq:endhour} reflect the requirement to deliver the reserve contracted at the start and end of the time period, respectively\footnote{We note that \eqref{eq:starthour} differs from the power-based start-of-hour ramping formulation proposed in \cite{nycander2021security}.}.
The total ramp requirement is determined by \cite{moradicapturing}
\begin{equation}
        \textstyle\sum_g PR_{g,d,t}  = 3\% \sum_l PL_{l,d,t} + 5\% \sum_w PW_{w,d,t}, \, \forall d,t
\end{equation}
The bus-specific demand patterns in each area and load shedding possibility are given by
\begin{subequations}
\begin{align}
    PL_{l,d,t} &= \textstyle\sum_{a,b} M^B_{b,a} M^L_{b,l} FL_{a,d,t} L_l  && \forall l,d,t \label{eq:load01} \\
    PS_{l,d,t} & \le 50\% PL_{l,d,t}  && \forall l,d,t\label{eq:lshed1}
\end{align}
\end{subequations}
here, $FL_{a,d,t}$ is instant load representative factors in each area. In constraint \eqref{eq:lshed1} instantaneous load shedding is bounded by a percentage of demand $PL_{l,d,t}$. The bus-specific dispatched wind power in each area is determined by
\begin{subequations}
\begin{align}
PW_{w,d,t} = \textstyle\sum_{a,b} [M^B_{b,a} M^W_{b,w} FW_{a,d,t} W_w] - PX_{w,d,t} \nonumber \\ \hspace{5cm} \forall w,d,t \\
0 \le PX_{w,d,t} \le \textstyle\sum_{a,b} M^B_{b,a} M^W_{b,w} FW_{a,d,t} W_w \hspace{0.25cm}\forall w,d,t   
\end{align}
\end{subequations}
where $FW_{a,d,t}$ and $PX_{w,d,t}$ represent instant wind representative factors in each area and wind curtailment.

\subsubsection{Energy Storage System}

Each ESS has grid-side power limit of $C^{cap}_s$, efficiencies $\eta^C_s$ (charging) and $\eta^D_s$ (discharging) and stored energy limit of $E^{cap}_s$. The evolution of stored energy levels is given by
\begin{multline} \label{storedElevel}
E_{s,d,t+1} =  E_{s,d,t} + \Delta_{d,t} \eta^C_s  \frac{PC_{s,d,t} + PC_{s,d,t+1}}{2} \\ 
- \frac{ \Delta_{d,t}}{\eta^D_s} \frac{ PD_{s,d,t} + PD_{s,d,t+1} }{2} \quad \forall s,d,t
\end{multline}
The charge and discharge power of ESS are constrained by
\begin{subequations} 
\label{eq:basicbattery}
\begin{align}
0 \le & \eta^C_s PC_{s,d,t}  \le C^{cap}_s && \forall s,d,t\\ 
0 \le & \frac{1}{\eta^D_s} PD_{s,d,t}  \le C^{cap}_s && \forall s,d,t\\
0 = & PC_{s,d,t} \times  PD_{s,d,t} && \forall s,d,t \label{eq:complementarity}
\end{align}
\end{subequations}
Here, the complementarity constraint \eqref{eq:complementarity} ensures that ESS can only charge \emph{or} discharge at $t$. We point out that in the PWL model, the complementarity constraint is only strictly enforced at time points $t$, but it is possible for the charging and discharging power to have (implied) nonzero values in between these time points. As this leads to a slightly conservative result (due to additional losses), we omit further constraints that would introduce additional nonlinearities. \eqref{eq:complementarity} is implemented using binary variables as:
\begin{subequations} 
\label{eq:linearization}
\begin{align}
\eta^C_s PC_{s,d,t}  \le U_{s,d,t} C^{max}_s && \forall s,d,t \\
\frac{1}{\eta^D_s} PD_{s,d,t}  \le (1-U_{s,d,t}) C^{max}_s && \forall s,d,t
\end{align}
\end{subequations}

\subsection{Reference Model}
In the reference model, stored energy levels are limited by
\begin{align}
    0 & \le  E_{s,d,t} \le E^{cap}_s && \forall s,d,t
\end{align}
Moreover, the $d$ indices refer to sequential days in the data set. Therefore, we impose the following boundary constraints on \emph{all} variables $X_{y,d,t}$ that have indices $d$, and $t$ (and an $X$-specific index $y$), to impose continuity between days and between the first and last day:
\begin{subequations} \label{eq:refboundary}
\begin{align}
    X_{y,d+1,0} &= X_{y,d,24} && \forall y,d \\
    X_{y,1,0} &= X_{y,|\mathcal{D}|,24} && \forall y
\end{align}
\end{subequations}
The objective function is to minimize the \emph{total planning cost}, consisting of the investment and operational parts.

\noindent\textbf{\emph{Reference Model:}}
\begin{align}
TPC^{\text{ref}} & = \min \left[ CI + \textstyle\sum_{d \in \mathcal{D}} CO_{d}  \right] \label{refmodel} \\
s.t: \quad \ 
& \text{constraints \eqref{eq:investmentcost}-\eqref{eq:refboundary}}, \nonumber \\
& \mathcal{D}=(\text{all days, sequentially}), \nonumber \\
&\mathcal{T}=(0,\dots,24), \nonumber \\
& \Delta_{d,t} = 1\text{hr} \hspace{1cm} \forall d,t \nonumber
\end{align}

\subsection{Representative Day Model}
When using RDs, the long-term ESS model must be adapted accordingly. We follow the ERD (Enhanced Representative Days) methodology \cite{gonzato2021long} with the extension to SLDs proposed in \cite{moradicapturing}. In this representation, the variables $E_{s,d,t}$ no longer represent the absolute state of the charge of the ESS, but its \emph{change} relative to the start of RD $d$. We define its total change in energy $E^{tot}_{s,d}$ and its minimum/maximum excursion during the day as follows:
\begin{subequations} \label{eq:ERDday}
\begin{align}
E^{tot}_{s,d} & = E_{s,d,24}, \ E_{s,d,0}=0 && \forall s,d \label{eq:ERDday_a}
\\ E^{low}_{s,d} &\le E_{s,d,t} \le E^{high}_{s,d} && \forall s,d,t \label{eq:ERDday_b}
\end{align}
\end{subequations}

In addition, each ESS is assigned a variable $LE_{s,sd}$ to track its state of charge at the beginning (and end) of each SLD. The association of RDs and SLDs is given by the matrix $D_{sd,d}$, with element values $\{0,1\}$ and the number of RD repetitions within an SLD is $n^{B}_{sd}$. The long-term stored energy is tracked for every SLD, using the following equations:
\begin{subequations} \label{eq:ERDyear}
\begin{align}
LE_{s,sd+1} & = LE_{s,sd} + n^B_{sd} \textstyle\sum_{d} D_{sd,d} E^{tot}_{s,d} \hspace{0.75cm} \forall s,sd \\
LE_{s,1} & = LE_{s,|\mathcal{SD}|}+ n^B_{|\mathcal{SD}|} \textstyle\sum_{d} D_{|\mathcal{SD}|,d} E^{tot}_{s,d} \hspace{0.4cm}  \forall s \\
0 & \le LE_{s,sd} + \textstyle\sum_{d} D_{sd,d} E^{low}_{s,d} \hspace{1.25cm} \forall s,sd \\
0 & \le LE_{s,sd} + \nonumber \\
& \textstyle\sum_{d} D_{sd,d} [(n^B_{sd}-1)E^{tot}_{s,d}+E^{low}_{s,d}]\hspace{0.5cm} \forall s,sd \\
E^{cap}_s & \ge LE_{s,sd} + \textstyle\sum_{d} D_{sd,d} E^{high}_{s,d} \hspace{1.11cm} \forall s,sd \\
E^{cap}_s & \ge LE_{s,sd} + \nonumber \\ 
& \textstyle\sum_{d} D_{sd,d} [ E^{high}_{i,d} + (n^B_{sd}-1) E^{tot}_{s,d} ] \hspace{0.4cm} \forall s,sd 
\end{align}
\end{subequations}
Combining all the above constraints results in the following RD model in PWL representation.

\noindent\textbf{\emph{Representative Day Model (RD):}}  \begin{align}
TPC^{\text{RD}} & = \min \left[ CI + \textstyle\sum_{d \in \mathcal{D}} \omega_d CO_{d} \right] \label{RDmodel} \\
s.t: \quad \ 
& \text{constraints \eqref{eq:investmentcost}-\eqref{eq:linearization}, \eqref{eq:ERDday}-\eqref{eq:ERDyear}}, \nonumber \\
& \mathcal{SD}=(1,\ldots,\text{\# of SLDs}), \nonumber \\ 
&\mathcal{D}=\{\text{RDs}\}, \nonumber \\
& \omega_d = \{\text{weight of each RD}\}, \hspace{0.8cm}  \forall d \nonumber \\
& \mathcal{T}=(0,\ldots, 24), \nonumber \\ 
&\Delta_{d,t} = 1\text{hr} \hspace{2.5cm}  \forall d,t \nonumber
\end{align}

\subsection{Representative Day $\&$ Time Point Model}
Finally, the RDTP model can be constructed from the previously presented constraints, using a set of RDs and unequal time points computed by \eqref{eq:opt_for_RTP_all}, either using equal or unequal allocation to of RTPs to RDs. The main difference with the RD model is that the set of time indices $\mathcal{T}_d$ now depends on the RD $d$. This also necessitates updating \eqref{eq:ERDday_a} to
\begin{equation} \label{eq:RDHday}
\begin{aligned}
E^{tot}_{s,d} &= E_{s,d,\max(\mathcal{T}_d)} \quad \quad \forall s,d 
\end{aligned}
\end{equation}
The final model is then given by:

\noindent\textbf{\emph{Representative Day \& Time Point Model (RDTP):}}  \begin{align}
TPC^{\text{RDTP}} & = \min \left[ CI + \textstyle\sum_{d \in \mathcal{D}} \omega_d CO_{d} \right] \label{RDTPmodel} \\
s.t: \quad \
& \text{constraints \eqref{eq:investmentcost}-\eqref{eq:linearization}, \eqref{eq:ERDday_b}, \eqref{eq:ERDyear}}, \eqref{eq:RDHday}, \nonumber \\
& \mathcal{SD}=(1,\ldots,\text{\# of SLDs}),\nonumber \\
&\mathcal{D}=\{\text{RDs}\}, \nonumber \\
& \omega_d = \{\text{weight of each RD}\},\qquad \hspace{1.56cm} \forall d \nonumber \\
& \mathcal{T}_d=(0,\ldots, {r_d-1}),\hspace{3.25cm} \forall d\nonumber \\
& \Delta_{d,t} = \text{variable, according to equation \eqref{durationRTP_2}} \quad \forall d,t \nonumber
\end{align}

\section{Performance analysis}

\subsection{Study System}
The effectiveness of the proposed co-planning model and representative period extraction methods was evaluated using the IEEE RTS 24-bus test system \cite{780914}. We considered seven areas with distinct load and RES generation patterns based on the Netherlands (buses 9, 11, 15, 16, 26), Belgium (buses 17, 18, 25), France (buses 1, 3, 4, 24), Germany (buses 6, 10, 12, 13), Denmark (buses 14, 19, 20), Sweden (buses 21, 22, 23), and Switzerland (buses 2, 5, 7, 8). Note that buses 25 and 26 with candidate wind farms are assumed to be expansion buses. All the data for this test system, including system topology, cost of existing and new candidate options, required parameters, along with data source references, are available in \cite{Modified24}. The one-year load and renewable energy generation data for each area are based on one year of data from 2019. The CPLEX solver in the GAMS environment \cite{General} was employed to solve the proposed MILP co-planning and optimization-based RTP extraction problems. Additionally, the multi-area clustering-based RD extraction algorithm was implemented in Matlab \cite{matlabb}, running on a PC with an Intel Xeon W-2223 CPU 3.60 GHz and 16 GB of RAM.

\subsection{Numerical Results}

The planning problem was solved using different model formulations. First, the reference model \textbf{Ref-PWL} \eqref{refmodel} was solved using all hours and days, with results shown in Table \ref{table1}. Four new transmission lines, four short-term ESS, and four long-term ESS, along with six wind farms were scheduled for installation and the required CPU time was $\approx 67$ hours, highlighting the importance of problem reduction. 

\begin{table}[t!]
\renewcommand{\arraystretch}{1.25}
\setlength{\tabcolsep}{5pt}
\caption{Results for the reference model (Ref-PWL)} 
\label{table1}
\centering
\resizebox{!}{1.7cm}{
\begin{tabular}{c|ccccc}
\hline \hline
\multirow{2}{*}{\textbf{\begin{tabular}[c]{@{}c@{}}Cost:\\  Value (×$10^6$\$):\end{tabular}}} & \multicolumn{1}{c|}{\textbf{TCO$^*$}} & \multicolumn{3}{c|}{\textbf{CI}} & \textbf{TPC} \\ \cline{2-6} 
 & \multicolumn{1}{c|}{2021.4} & \multicolumn{3}{c|}{776.4} & 2797.8 \\ \hline
\multirow{2}{*}{\textbf{Planning Option:}} & \multicolumn{1}{c|}{\multirow{2}{*}{\textbf{\begin{tabular}[c]{@{}c@{}}Transmission\\ Line\end{tabular}}}} & \multicolumn{3}{c|}{\textbf{ESS (bus)}} & \multirow{2}{*}{\textbf{\begin{tabular}[c]{@{}c@{}}Wind Farm\\ (bus)\end{tabular}}} \\ \cline{3-5}
 & \multicolumn{1}{c|}{} & \multicolumn{2}{c|}{\textbf{Short-term}} & \multicolumn{1}{c|}{\textbf{Long-term}} &  \\ \hline
\textbf{Location:} & \multicolumn{1}{c|}{\begin{tabular}[c]{@{}c@{}}(7-8), (14-16),\\ (19-25), (16-26)\end{tabular}} & \multicolumn{2}{c|}{2, 3, 16, 25} & \multicolumn{1}{c|}{8, 11, 19, 23} & \multicolumn{1}{c} {\begin{tabular}[c]{@{}c@{}} 3, 5, 6, 14,\\ 25, 26 \end{tabular}} \\ \hline
\textbf{Load Shedding (GWh)}  & \multicolumn{5}{c}{4.92} \\ \hline
\textbf{CPU Time (Sec):} & \multicolumn{5}{c}{$242,825$} \\ \hline \hline
 \multicolumn{1}{c}{*: Total Cost of Operation}
\end{tabular}%
}
\end{table}

To facilitate a comparison between PWL and PWC models, an equivalent PWC reference model was constructed, by reformulating equations \eqref{LScost}, \eqref{eq:costfunction}-\eqref{eq:linconstraint3}, \eqref{storedElevel} and \eqref{eq:refboundary} to use 24 constant power values. The investment decisions for the reference PWC model were identical, and the other costs were comparable with the PWL model (within the optimality gap).  
However, the advantages of the PWL approach become evident when approximating the model in reduced space cases with RDs and RTPs.  

Models with representative days and/or time periods were compared to their respective reference models on the basis of investment cost, operation cost, and total cost. The error metric used to analyze the effectiveness of the proposed method is
\begin{equation}\label{eq:error}
\textrm{Error}=\frac{f_c^*(\widehat{v})-f_c^*(v^*)}{f_c^*(v^*)}.
\end{equation}
Here $v^*$ represents the decision variables for the reference case, while $\widehat{v}$ denotes the fixed decision variables obtained from the reduced space models. $f_c^*$ signifies the desired cost type,  assessed across all days and hours (non-reduced).

\begin{figure}[!t]
\centering
\includegraphics[scale=0.415]{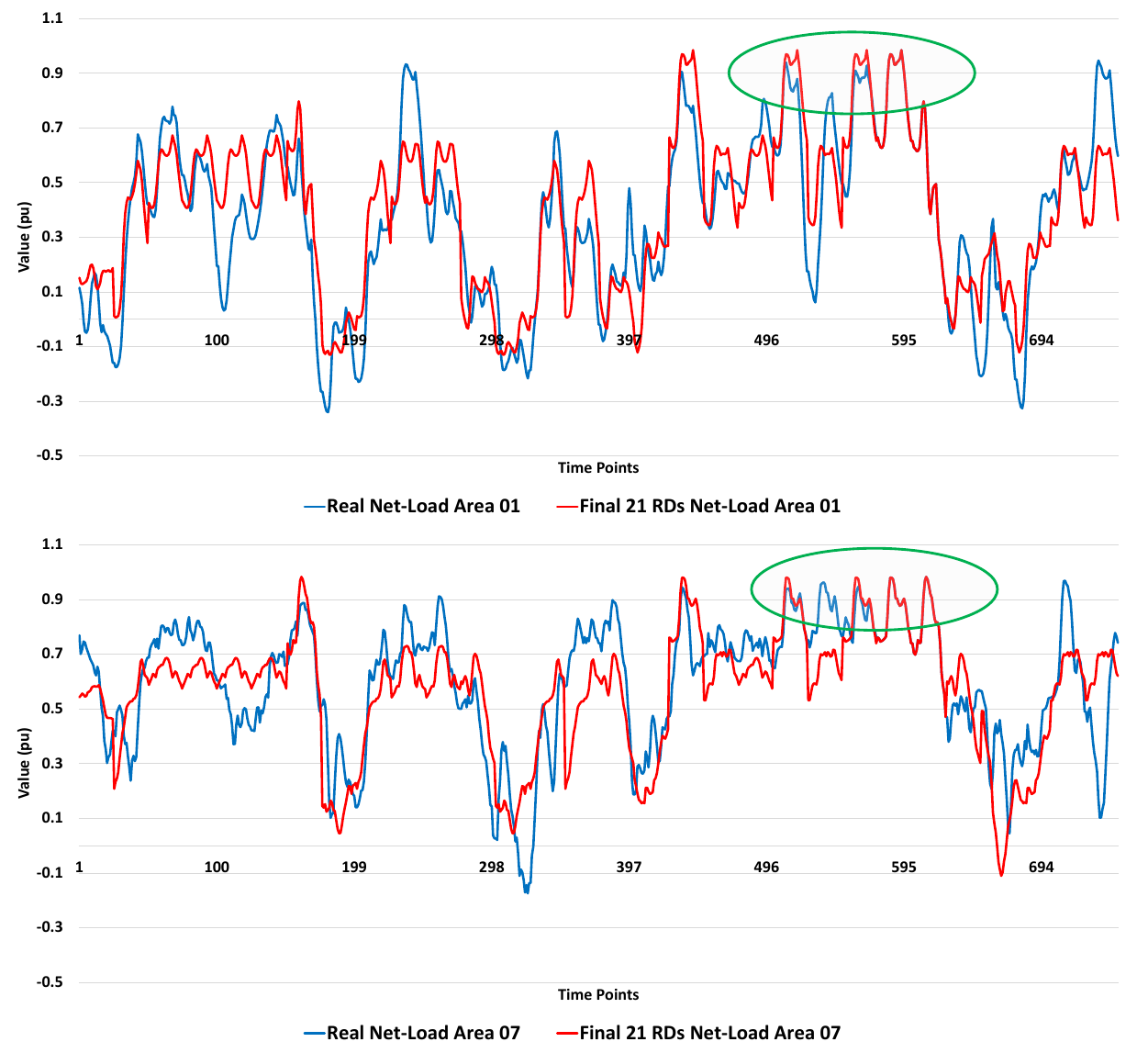}
\caption{Comparison of original net load and reconstructed net load based on RDs, for two areas (top and bottom), for one month with hourly resolution. The ability to capture net load peaks using extreme day preservation is highlighted}
\label{fig:002}
\end{figure}

\begin{table}[b!]
\renewcommand{\arraystretch}{1.35}
\caption{Error and CPU time comparison between cases RD-noEx \& RD} 
\label{table2}
\centering
\resizebox{!}{1.1cm}{
\begin{tabular}{c|c|c|c|c}
\hline \hline \multirow{2}{*} { \textbf{Case} } & \multicolumn{3}{c|}{ \textbf{Error (\%)} } & \multirow{2}{*}{ \textbf{CPU Time (sec):} } \\
\cline { 2 - 4 } & \textbf{Operation} & \textbf{Investment} & \textbf{Total} & \\
\hline $\textbf{RD-noEx}$ & $6.214$ & $-5.312$ & $2.908$ & 414 \\
\hline $\textbf{RD}$ & $-0.041$ & $0.945$ & $0.234$ & 637 \\
\hline \hline
\end{tabular}}
\end{table}
In cases \textbf{RD-noEx} and \textbf{RD}, the PWL RD model \eqref{RDmodel} was evaluated with 21 extracted RDs, ignoring and considering extreme value days, respectively. The number of resulting SLDs was 271 for \textbf{RD-noEx}, and 267 for \textbf{RD}. The results in Table~\ref{table2} illustrate the significance of capturing extreme value days. The error in total planning cost reduced dramatically from nearly 3\% to 0.23\%, resulting from reductions in investment and operation cost errors. Fig.~\ref{fig:002} highlights the ability of the extreme-day sensitive RD selection process to capture net load peaks. The use of RDs reduced CPU time by up to 99.7\%. For subsequent results, the extreme day selection was used.

\begin{table}[b!]
\renewcommand{\arraystretch}{1.15}
\caption{Equal and adaptive number of RTPs comparison within each RD} 
\label{table03}
\centering
\resizebox{!}{2.70cm}{
\begin{tabular}{c|c|c|c|c|c}
\hline \hline \multirow{2}{*}{\textbf{RD:}} & \multicolumn{2}{c|}{ \textbf{\#of Extracted RTP} } & \multirow{2}{*}{\textbf{RD:}} & \multicolumn{2}{c}{ \textbf{\#of Extracted RTP}} \\
\cline { 2-3 } \cline { 5-6 } & \textbf{Equal:} & \textbf{Adaptive:} && \textbf{Equal:} & \textbf{Adaptive:} \\
\hline $\mathbf{1}$ & 10 & 09 & $\mathbf{12}$ & 10 & 12 \\
\hline $\mathbf{2}$ & 10 & 10 & $\mathbf{13}$ & 10 & 11 \\
\hline $\mathbf{3}$ & 10 & 10 & $\mathbf{14}$ & 10 & 10 \\
\hline $\mathbf{4}$ & 10 & 09 & $\mathbf{15}$ & 10 & 11 \\
\hline $\mathbf{5}$ & 10 & 10 & $\mathbf{16}$ & 10 & 09 \\
\hline $\mathbf{6}$ & 10 & 10 & $\mathbf{17}$ & 10 & 08 \\
\hline $\mathbf{7}$ & 10 & 11 & $\mathbf{18}$ & 10 & 09 \\
\hline $\mathbf{8}$ & 10 & 11 & $\mathbf{19}$ & 10 & 09 \\
\hline $\mathbf{9}$ & 10 & 11 & $\mathbf{20}$ & 10 & 09 \\
\hline $\mathbf{10}$ & 10 & 13 & $\mathbf{21}$ & 10 & 09 \\
\hline $\mathbf{11}$ & 10 & 10 & \textbf{TNRTP$^*$} & $210$ & $210$ \\
\hline  \multicolumn{4}{c|}{ \textbf{ATAE$^{**}$(pu)} } & \textbf{1.602} & \textbf{1.578} \\ \hline \hline \multicolumn{6}{c}{*: Total Number of RTPs, \  **: Average of Total Absolute Error} 
\end{tabular}}
\end{table} 

\begin{figure}[!t]
\centering
\includegraphics[scale=0.3483]{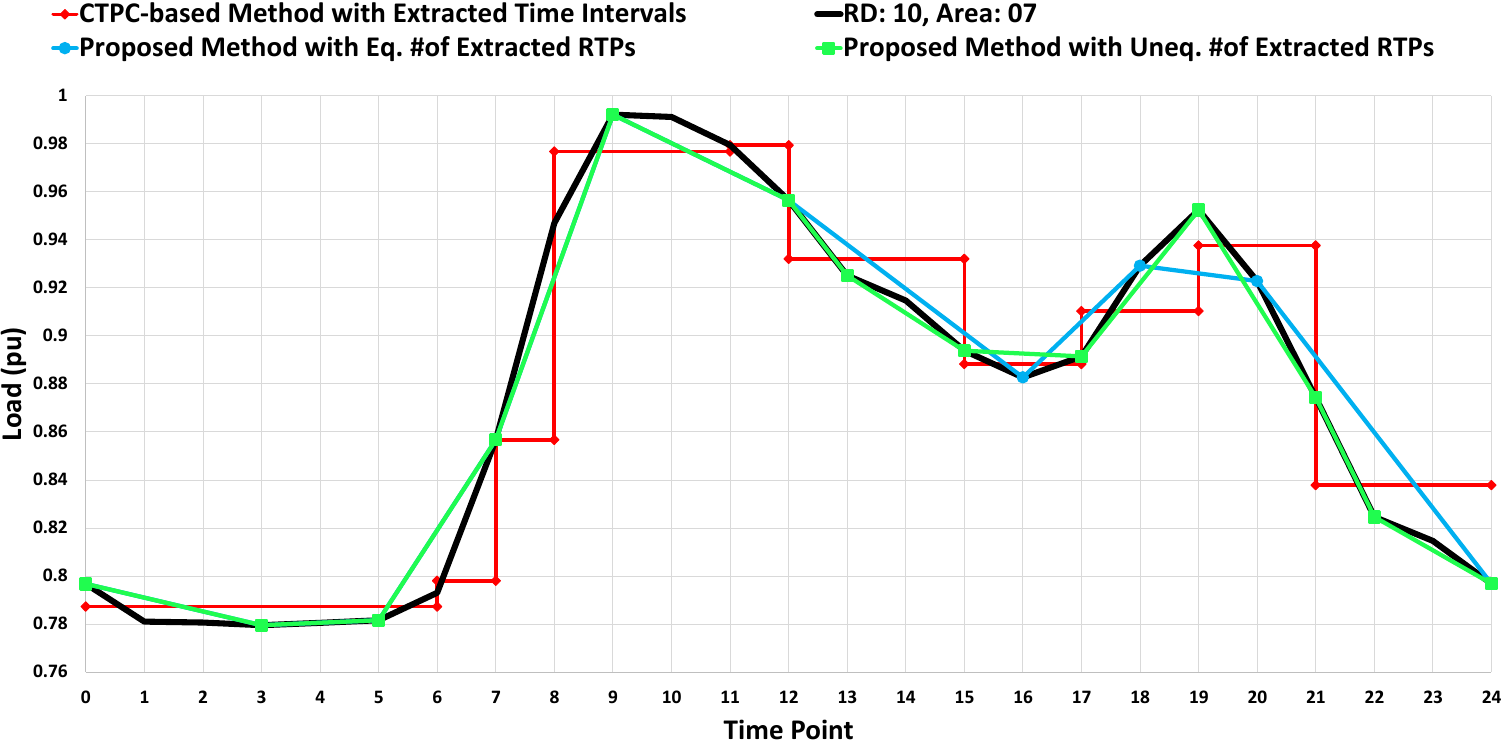}
\caption{Comparison of extraction of equal and unequal number of RTPs for PWL models, alongside the CTPC method for PWC models} 
\label{fig:003}
\end{figure}

The efficacy of the \emph{RDTP model} \eqref{RDTPmodel} was investigated next. The benefit of an adaptive allocation of RTPs across RDs can be illustrated in two other ways. First, as presented in Table \ref{table03}, after executing the proposed optimization-based method \eqref{eq:opt_for_RTP_all}-\eqref{durationRTP_2} to extract 10 RTPs within each of the 21 extracted RDs, for a total of 210 RTPs, the adaptive selection of numbers of RTPs resulted in an average total absolute error (i.e., \eqref{eq:opt_for_RTP_a}) of 1.578 pu, compared to a value of 1.602 pu for the case where each RD was allocated exactly 10 RTPs. Additionally, Fig. \ref{fig:003}  provides an illustrative example, comparing both methods of extracting equal and unequal numbers of RTPs with the CTPC method, for load data of area 07 in RD 10. The adaptive method assigned 13 RTPs to this relatively variable day, allowing it to capture more details. Moreover, both PWL-based methods are able to better approximate ramps, even when larger time steps occur between RTPs.
\begin{table}[b!]
\renewcommand{\arraystretch}{1.36}
\caption{Error and CPU time comparison between cases RD-CTPC, RDTP-eq \& RDTP-ad} 
\label{table04}
\centering
\resizebox{!}{1.32cm}{
\begin{tabular}{c|c|c|c|c}
\hline \hline \multirow{2}{*} { \textbf{Case} } & \multicolumn{3}{c|}{ \textbf{Error (\%) }} & \multirow{2}{*}{ \textbf{CPU Time (sec):} } \\
\cline { 2 - 4 } & \textbf{Operation} & \textbf{Investment} & \textbf{Total} & \\
\hline $\textbf{RD-CTPC}$ & $-0.210$ & $2.571$ & $0.562$ & 84 \\
\hline $\textbf{RDTP-eq}$ & $-0.089$ & $1.712$ & $0.411$ & 97 \\
\hline $\textbf{RDTP-ad}$ & $-0.075$ & $1.449$ & $0.348$ & 95 \\
\hline \hline
\end{tabular}}
\end{table}

The first case, \textbf{RD-CTPC}, is based on a PWC formulation. 10 representative time intervals were extracted from each of the 21 extracted RDs using the CTPC method \cite{pineda2018chronological}. Planning errors were calculated using \eqref{eq:error} by comparison with the PWC reference model and the obtained results are presented in Table~\ref{table04}. A significant reduction in CPU time is observed due to the use of RDs.

In cases \textbf{RDTP-eq} and \textbf{RDTP-ad}, the proposed optimization-based method \eqref{eq:opt_for_RTP_a}-\eqref{durationRTP_2} was utilized to extract 10 RTPs, with equal and unequal RTPs within each of the 21 RDs, respectively. The results were compared to Ref-PWL in Table~\ref{table04}. Both PWL-based methods outperform the PWC-based approach with CTPC, and the adaptive allocation of RTPs across RDs yielded a slight further improvement. Compared to the RD methods, a speedup of approximately 6 times was obtained, at a very minor reduction in accuracy (comparing \textbf{RD} with \textbf{RDTP-ad}). Compared to the reference case \textbf{Ref-PWL}, the TPC and CI in case \textbf{RDTP-ad} increased by $0.348\%$ and $1.449\%$, respectively, and the TCO decreased by $0.075\%$, as also presented in Table \ref{table04}. Similar to the \textbf{RDTP-eq} method, the CPU time was significantly reduced for both RDTP methods. Jointly, the reduced error and large computational savings confirm the effectiveness of the proposed \textbf{RDTP-ad} method in balancing complexity and accuracy.

Table \ref{table5} shows the investment decision variables, along with TCO, CI, and TPC for case \textbf{RDTP-ad}. The constructed transmission lines, as well as the installed ESS and wind farm locations, were the same for both cases \textbf{Ref-PWL} and \textbf{RDTP-ad}. The reason for the different costs is attributed to the installed capacities of the ESS and wind farms. In case \textbf{RDTP-ad}, more capacities were installed to cover extreme values in the data, which led to a higher CI and lower TCO. Consequently, load shedding was also slightly less in case \textbf{RDTP-ad}. 
\begin{table}[t!]
\renewcommand{\arraystretch}{1.25}
\setlength{\tabcolsep}{5pt}
\caption{Results for case RDTP-ad} 
\label{table5}
\centering
\resizebox{!}{1.55cm}{
\begin{tabular}{c|ccccc}
\hline \hline
\multirow{2}{*}{\textbf{\begin{tabular}[c]{@{}c@{}}Cost:\\  Value (×$10^6$\$):\end{tabular}}} & \multicolumn{1}{c|}{\textbf{TCO}} & \multicolumn{3}{c|}{\textbf{CI}} & \textbf{TPC} \\ \cline{2-6} 
 & \multicolumn{1}{c|}{2019.88} & \multicolumn{3}{c|}{787.65} & 2807.53 \\ \hline
\multirow{2}{*}{\textbf{Planning Option:}} & \multicolumn{1}{c|}{\multirow{2}{*}{\textbf{\begin{tabular}[c]{@{}c@{}}Transmission\\ Line\end{tabular}}}} & \multicolumn{3}{c|}{\textbf{ESS (bus)}} & \multirow{2}{*}{\textbf{\begin{tabular}[c]{@{}c@{}}Wind Farm\\ (bus)\end{tabular}}} \\ \cline{3-5}
 & \multicolumn{1}{c|}{} & \multicolumn{2}{c|}{\textbf{Short-term}} & \multicolumn{1}{c|}{\textbf{Long-term}} &  \\ \hline
\textbf{Location:} & \multicolumn{1}{c|}{\begin{tabular}[c]{@{}c@{}}(7-8), (14-16),\\ (19-25), (16-26)\end{tabular}} & \multicolumn{2}{c|}{2, 3, 16, 25} & \multicolumn{1}{c|}{8, 11, 19, 23} & \multicolumn{1}{c} {\begin{tabular}[c]{@{}c@{}} 3, 5, 6, 14,\\ 25, 26 \end{tabular}} \\ \hline
\textbf{Load Shedding (GWh)}  & \multicolumn{5}{c}{4.71} \\ \hline
\textbf{CPU Time (Sec):} & \multicolumn{5}{c}{95} \\ \hline \hline
\end{tabular}%
}
\end{table}
\section{Conclusion}
This paper proposed a hybrid multi-area piecewise linear (PWL) adapted method for capturing interday and intraday chronology, considering extreme values. This was achieved by extracting representative days (RDs) and time points (RTPs) within each RD, addressing a complex co-planning problem involving transmission lines, energy storage systems, and wind farms. To enhance the representation of intraday chronology, an optimization-based RTP extraction approach was introduced that adaptively extracts unequal numbers of RTPs within each RD. The effectiveness of the proposed model with PWL OPF formulations was evaluated using six different cases. The importance of capturing extreme values was demonstrated. Moreover, the ability of the adaptive method to assign more RTPs to RDs with higher hourly variation facilitates the accurate representation of intraday chronology. In the case study, the combination of these approaches resulted in a time saving of $>99.9\%$, accompanied by a total planning cost error of only 0.348\%.

Future studies might investigate dealing with a decade of input data for larger areas and additional renewable energy sources, such as solar, can be explored to further validate and expand our findings. Moreover, for optimal deployment of the method across a range of practical scenarios, it will be important to establish rules of thumb on how many RTPs to identify across which number of RDs, for a given computational budget.


\end{document}